\documentclass[aps,pra,reprint,amsmath,amsfonts,amssymb,superscriptaddress]{revtex4-1}
\usepackage{graphicx,float,calc}
\usepackage{color,bm}
\usepackage{ulem}
\usepackage{braket}
\usepackage[colorlinks,urlcolor=blue,citecolor=blue,linkcolor=blue]{hyperref}
\begin{document}
\title{Localization and topological transitions in non-Hermitian quasiperiodic lattices}
\author{Ling-Zhi Tang}
\affiliation{Guangdong Provincial Key Laboratory of Quantum Engineering and Quantum Materials, School of Physics and Telecommunication Engineering, South China Normal University, Guangzhou 510006, China}
\author{Guo-Qing Zhang}\thanks{zhangptnoone@m.scnu.edu.cn}
\affiliation{Guangdong Provincial Key Laboratory of Quantum Engineering and Quantum Materials, School of Physics and Telecommunication Engineering, South China Normal University, Guangzhou 510006, China}
\affiliation{Guangdong-Hong Kong Joint Laboratory of Quantum Matter, Frontier Research Institute for Physics, South China Normal University, Guangzhou 510006, China}
\author{Ling-Feng Zhang}
\affiliation{Guangdong Provincial Key Laboratory of Quantum Engineering and Quantum Materials, School of Physics and Telecommunication Engineering, South China Normal University, Guangzhou 510006, China}
\author{Dan-Wei Zhang}\thanks{danweizhang@m.scnu.edu.cn}
\affiliation{Guangdong Provincial Key Laboratory of Quantum Engineering and Quantum Materials, School of Physics and Telecommunication Engineering, South China Normal University, Guangzhou 510006, China}
\affiliation{Guangdong-Hong Kong Joint Laboratory of Quantum Matter, Frontier Research Institute for Physics, South China Normal University, Guangzhou 510006, China}

\begin{abstract}
We investigate the localization and topological transitions in a one-dimensional (interacting) non-Hermitian quasiperiodic lattice, which is described by a generalized Aubry-Andr\'{e}-Harper model with irrational modulations in the off-diagonal hopping and on-site potential and with non-Hermiticities from the nonreciprocal hopping and complex potential phase. For noninteracting cases, we reveal that the nonreciprocal hopping (the complex potential phase) can enlarge the delocalization (localization) region in the phase diagrams spanned by two quasiperiodic modulation strengths. We show that the localization transition are always accompanied by a topological phase transition characterized the winding numbers of eigenenergies in three different non-Hermitian cases. Moreover, we find that a real-complex eigenenergy transition in the energy spectrum coincides with (occurs before) these two phase transitions in the nonreciprocal (complex potential) case, while the real-complex transition is absent under the coexistence of the two non-Hermiticities. For interacting spinless fermions, we demonstrate that the extended phase and the many-body localized phase can be identified by the entanglement entropy of eigenstates and the level statistics of complex eigenenergies. By making the critical scaling analysis, we further show that the many-body localization transition coincides with the real-complex transition and occurs before the topological transition in the nonreciprocal case, which are absent in the complex phase case.
\end{abstract}

\date{\today}

\maketitle

\section{Introduction}

\begin{figure}[tb]
	\centering
	\includegraphics[width=0.45\textwidth]{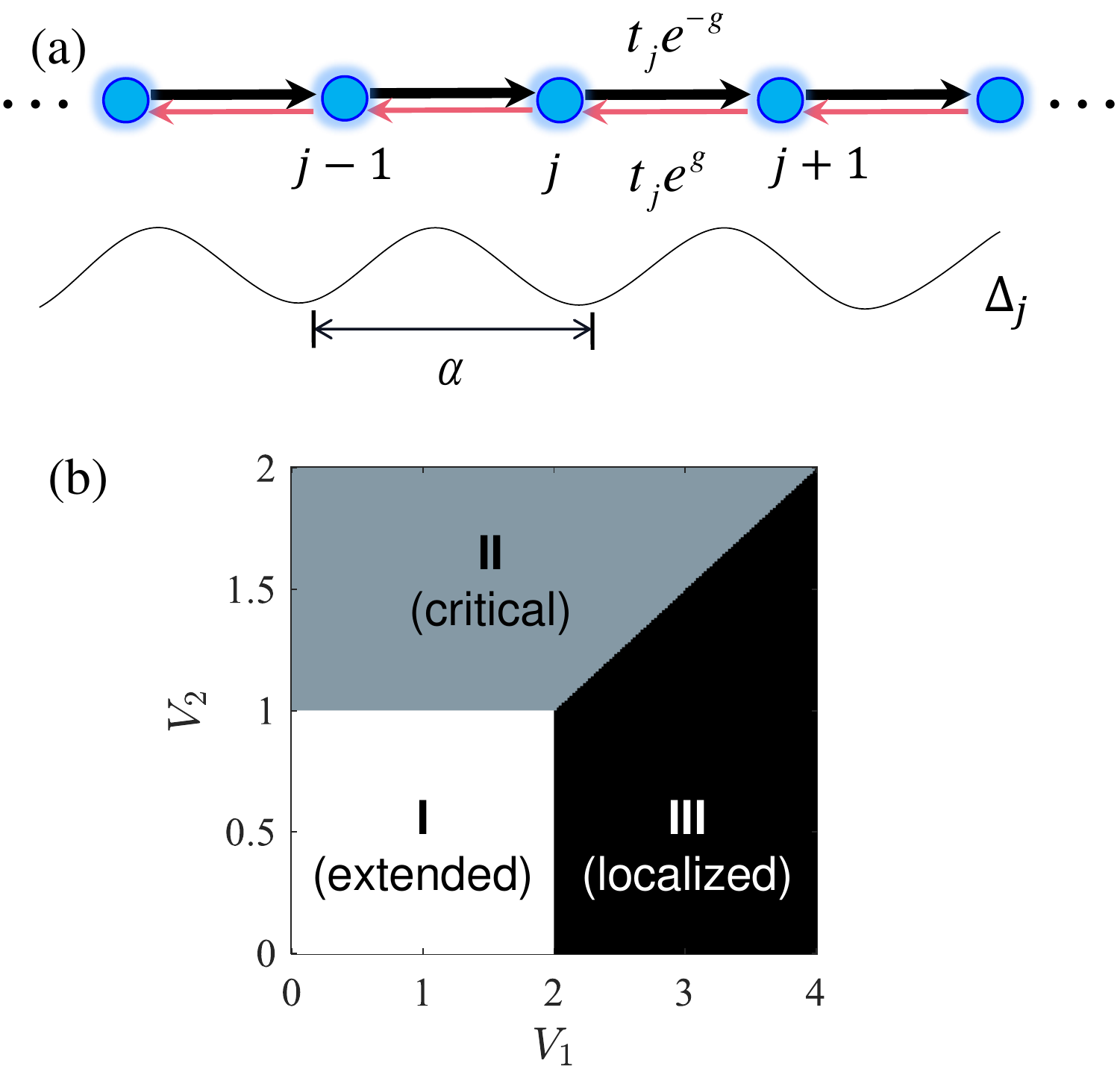}
	\caption{(Color online) (a) Sketch of the non-Hermitian generalized AAH model.
		$t_je^{g}$ $(t_je^{-g})$ represents the nonreciprocal left (right) hopping, $\Delta_j$ corresponds to the on-site potential and $\alpha=(\sqrt{5}-1)/2$ is the quasiperiodic modulation. (b) Phase diagram in the Hermitian and noninteracting limits with the phase boundary determined by $V_1=2\mathrm{max}(t, V_2)$ and $V_2=t$ \cite{Thouless1983,Chang1997,FLLiu2014,YCWang2019}, where $t=1$ is set as the energy unit. The phase diagram consists three regions $\mathrm{I}, \mathrm{II}, \mathrm{III}$\ for the extended, critical, and localized phases, respectively.
	}\label{fig1}
\end{figure}

The Aubry-Andr\'{e}-Harper (AAH) model~\cite{Harper1955,Aubry1980} serves as an important platform to study the Anderson localization~\cite{Anderson1958} and topological states of matter~\cite{XLQi2011,DWZhang2018} in one-dimensional (1D) quasiperiodic lattices~\cite{Aubry1980,Harper1955}. It has been found that the localization transition for all eigenstates and the topological pumping can be achieved by the quasiperiodic potential modulation in the AAH model. Recently, the AAH model and its various generalizations have received lots of attention~\cite{Kraus2012a,Verbin2013,Madsen2013,LJLang2012,Ganeshan2013,Lado2019,Vidal1999,Vidal2001} due to their realizations in some artificial systems, such as photonic crystals~\cite{DalNegro2003,Lahini2009,Kraus2012} and ultracold atoms~\cite{Modugno2010, Roati2008a}.
One of the generalized AAH models is by imposing incommensurate modulation on both on-site potential and off-diagonal hopping amplitude \cite{Thouless1983,Chang1997,FLLiu2014,YCWang2019}. Such a generalized AAH model with two quasiperiodic modulation parameters has a critical phase lying between the extended and localized phases in the phase diagram [see Fig. (\ref{fig1})]~\cite{Thouless1983,Chang1997,FLLiu2014,YCWang2019}. By imposing inter-particle interactions into the AAH model, one can explore the physics of many-body localization (MBL), which is the interacting counterpart of Anderson localization~\cite{Lueschen2017,Kohlert2019,Bordia2016}.

On the other hand, growing effort has been taken to explore the localization and topological properties in various non-Hermitian systems ~\cite{Hatano1996,Hatano1997,Bender1998,Bender2007,ZGong2018,Longhi2019a,Longhi2019b,HJiang2019,DWZhang2019,XWLuo2019,LZTang2020,HLiu2020,QBZeng2020,Hamazaki2019,LJZhai2020,QBZeng2020,TLiu2020b,DWZhang2020,TLiu2020,ZXu2020,HQWang2018,FSong2019,TLiu2019,Ghatak2019,TSDeng2019,Ezawa2019,CHLiu2019,Kawabata2019,LJin2019,Longhi2019,JHou2019,Yoshida2019,Borgnia2019,Yamamoto2019,Herviou2019,Hamazaki2019,SYao2018,Kunst2018,Alvarez2018,SYao2018b,HShen2018,El-Ganainy2018,CYin2018,YChen2018,YXiong2018,LZhou2018,JQCai2018,MPan2018,Yuce20182,CHYin2018,Yuce2018,HJiang2018,LWZhou2018,RWang2018,Longhi2017,LFeng2017,Leykam2017,Jin2017,Lee2016,Fring2016,Longhi2015b,Longhi2015a,Malzard2015,Zeuner2015,XWang2015,Longhi2014,Lee2014a,Lee2014b,BZhu2014,SDLiang2013,Esaki2011,YHu2011,Rudner2009}, such as the non-Hermitian generalizations of AAH model.
In these systems, the non-Hermiticities can come from the nonreciprocal hopping terms or complex on-site potentials. Remarkably,
for noninteracting AAH model in the presence of one of these two non-Hermiticities, it has been theoretically found that the Anderson localization transition coincides with topological transitions~\cite{Longhi2019a,HJiang2019}. The MBL of hard-core bosons in 1D non-Hermitian lattices with random \cite{Hamazaki2019} and quasiperiodic \cite{LJZhai2020} modulations on the on-site potential has been studied, respectively. It was shown that the MBL transition coincide with a real-complex eigenenergy transition (and a topological transition) in non-Hermitian many-body bosonic systems \cite{Hamazaki2019,LJZhai2020}. However, the non-Hermitian localization (MBL) and topological properties in the generalized AAH model with both quasiperiodic diagonal and off-diagonal modulations are yet to be studied. In particular, different non-Hermitian effects on the localization and topological transitions in the single-particle regime and the non-Hermitian MBL for interacting fermions in the quasiperiodic lattices are largely unexplored.

In this paper, we study the localization and topological transitions in (interacting) non-Hermitian quasiperiodic lattices of spinless fermions, which is described by the generalized AAH model with irrational modulations in the off-diagonal hopping and on-site potential and with non-Hermiticities arising from the nonreciprocal hopping and the complex potential phase. For noninteracting cases, we reveal that the nonreciprocal hopping (the complex potential phase) enlarges the delocalization (localization) region in the phase diagrams spanned by two  quasiperiodic modulation strengths. We show that the localization transition are always accompanied by a topological phase transition characterized by the winding numbers of eigenenergies in three different non-Hermitian cases. Moreover, we find that the real-complex eigenenergy transition in the energy spectrum coincides with (occurs before) these two phase transitions in the nonreciprocal (complex potential) case, while the real-complex transition is absent under the coexistence of the two non-Hermiticities. For interacting spinless fermions, we demonstrate that the extended phase and the MBL phase can be identified by the entanglement entropy (EE) of eigenstates and the level statistics of complex eigenenergies. By making the critical scaling analysis, we further show that the MBL transition coincides with the real-complex transition and occurs before the topological transition in the nonreciprocal case, which are absent in the complex phase case. Our work may stimulate further exploration of the interplay among localization, topology, and non-Hermiticity in quantum many-body systems.

The rest of this paper is organized as follows. We first propose the non-Hermitian generalized AAH model of interacting fermions with the nonreciprocal hopping and complex potential phase in Sec.~\ref{sec2}. Section~\ref{sec3} is devoted to reveal the single-particle localization and topological transitions under different non-Hermitian circumstances. In Sec. \ref{sec4}, we study the non-Hermitian MBL phase associated with the localization and topological transitions in the interacting quasiperiodic lattice. A brief summary is finally presented in Sec.~\ref{sec5}.

\section{\label{sec2}Model Hamiltonian}

We begin by considering a non-Hermitian generalized AAH quasiperiodic lattice with nonreciprocal hopping terms and complex potentials, as shown in Fig. \ref{fig1}(a). For spinless interacting fermions, the lattice system can be described by the following tight-binding Hamiltonian
\begin{equation}\label{H}
\begin{split}
H=&\sum_j [t_{j}(e^{-g+i\theta_g/L} c^\dagger_{j+1} c_j+e^{g-i\theta_g/L} c^\dagger_j c_{j+1})\\
&+\Delta_j c^\dagger_j c_{j}+Un_jn_{j+1}],
\end{split}
\end{equation}
where $L$ is the lattice length, $c_j$ $(c_j^\dagger)$ denotes the fermion annihilation (creation) operator on site $j$, and $U$ is the interaction strength. The hopping strength $t_{j}$ and the on-site potential $\Delta_j$ are given by
\begin{equation}\label{H2}
\begin{split}
t_{j}&=t+V_2\cos[2\pi (j+1/2)\alpha+\theta_h/L],\\
\Delta_j&=V_1\cos(2\pi j\alpha+\theta_h/L+ih),\\
\end{split}
\end{equation}
where $V_1$ and $V_2$ denote the modulation amplitudes of the on-site potential and the off-diagonal hopping, respectively. The parameter $\alpha$ is chosen as an irrational number to ensure the incommensurate potential. Here $\theta_g$ and $\theta_h$ are additional modulation phases varying from $0$ to $2\pi$ used solely for defining the winding numbers [see Eq. (\ref{winding number})]. We take $\theta_g=\theta_h=0$ in Eqs. (\ref{H}) and (\ref{H2}) unless calculating the winding numbers. The non-Hermiticities in this model is tuned by the nonreciprocal strength $g$ and the complex phase $h$. Hereafter, we set $t=1$ as the energy unit, choose the irrational number $\alpha=(\sqrt{5}-1)/2$ as the golden ratio and consider the periodic boundary condition in our numerical calculations.

In the noninteracting case $U=0$ and in the Hermitian limit with $g=h=0$, our model reduces to the generalized AAH model whose localization properties have been studied in Refs.~\cite{Thouless1983,Chang1997,FLLiu2014,YCWang2019}. In this case,  the phase diagram of the model is shown in Fig. \ref{fig1}(b). It has been found that all the eigenstates of the system are localized when $V_1>2 \mathrm{max}(t,V_2)$, the bulk eigenstates become extended when $\{V_1,2V_2\}<2t$, and the bulk eigenstates are critical in the rest of the parameter region. In the following, we investigate the non-Hermitian effects on the localization (MBL) and topological properties in the model in the presence of different non-Hermiticities and interactions of spinless fermions.

\begin{figure*}[tb]
	\centering
	\includegraphics[width=0.9\textwidth]{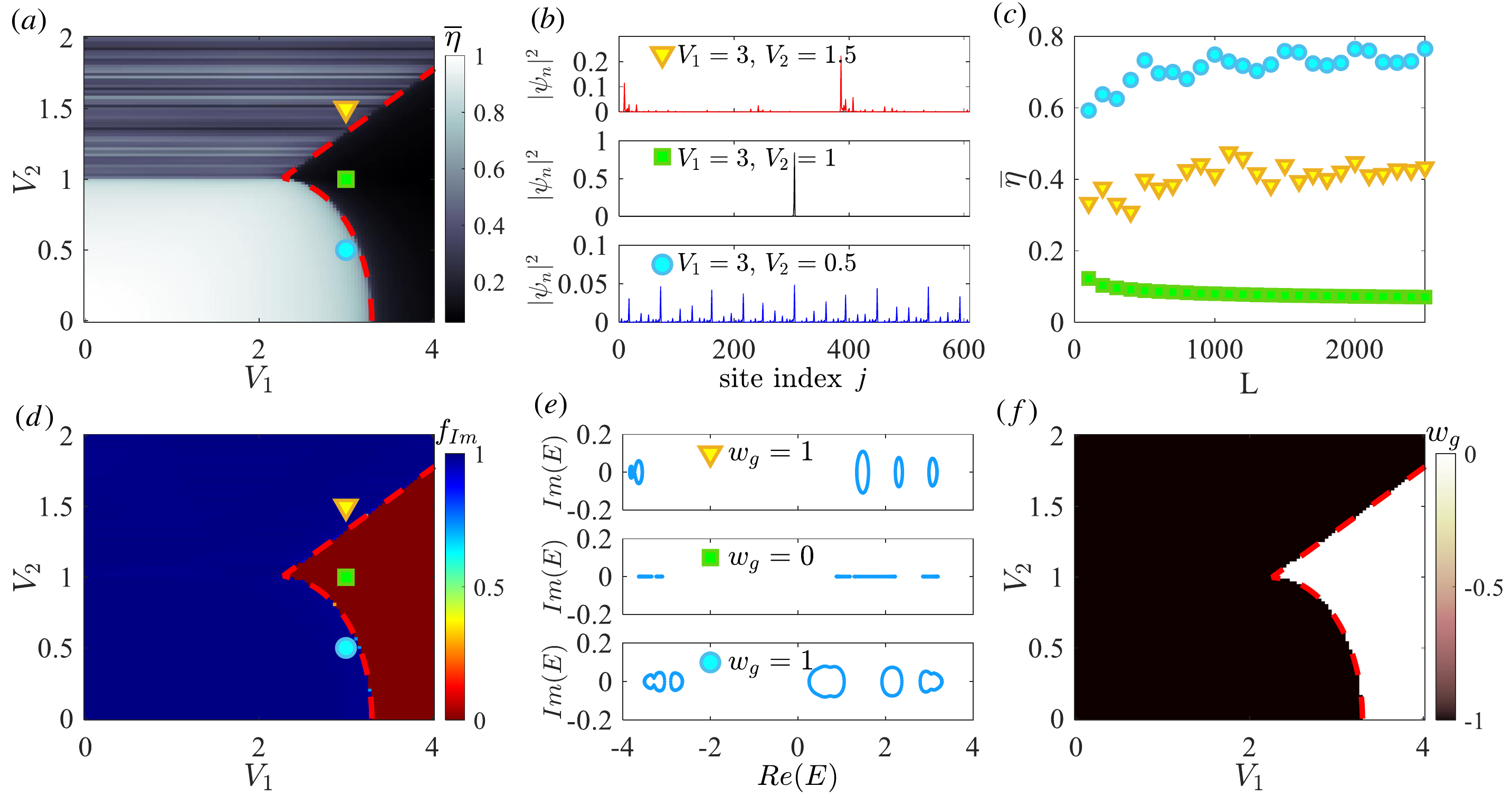}
	\caption{(Color online) The generalized AAH model with nonreciprocal hopping for the system size $L=610$. (a) The averaged FD $\overline{\eta}$ plotted in the parameter space ($V_1$,$V_2$) with the red dashed curve being the phase boundary determined by Eq. (\ref{boundary}). (b) The density distribution $|\psi_n(j)|^2$ of the ground state ($n=1$). (c) The scaling of $\overline{\eta}$ with the system size $L$ for the same $V_1$ and $V_2$ as in (b). (d) $f_{\text{Im}}$ as a function of $V_1$ and $V_2$. (e) Complex energy spectra for three different phases with the same $V_1$ and $V_2$ in (b). (f) The winding number $w_g$ as a function of $V_1$ and $V_2$. Other parameters are $g=0.5$ and $h=0$.
	}\label{fig2}
\end{figure*}

\section{\label{sec3} results in noninteracting cases}
In this section, we emphasize in the noninteracting case with $U=0$ and numerically study the localization and topological transitions in the non-Hermitian quasiperiodic lattice. In order to reveal different non-Hermitian effects on the phase transitions, we consider the nonreciprocal hopping, the complex potential, and the coexistence of both non-Hermiticities in the system, respectively. In our numerical simulations, we set the lattice size $L=610$, which is large enough for the quasiperiodic lattice in the noninteracting case.

\subsection{Nonreciprocal hopping case}

\begin{figure*}[tb]
	\centering
	\includegraphics[width=0.9\textwidth]{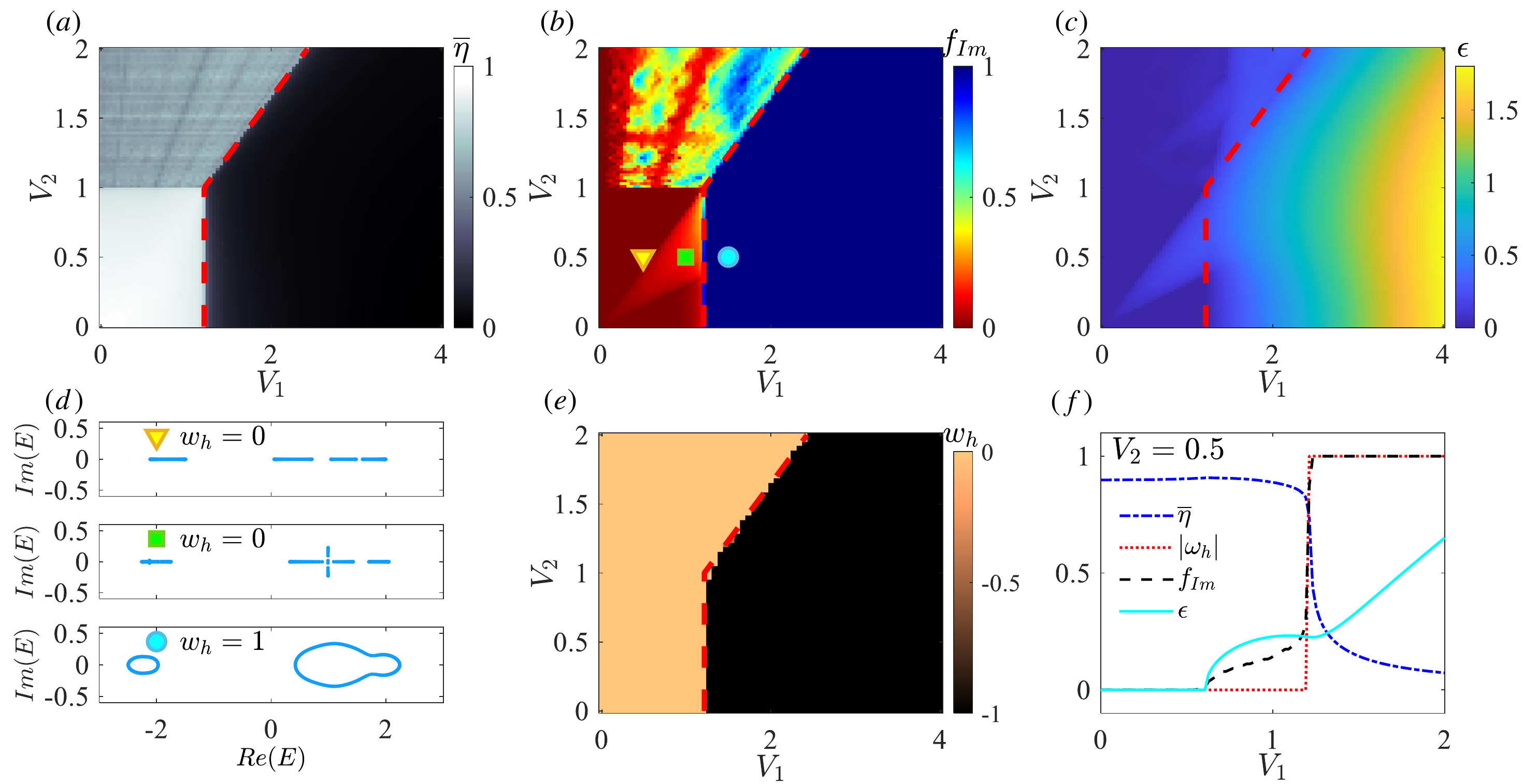}
	\caption{(Color online) The generalized AAH model with complex potential phase for the system size $L=610$. (a,b,c,e) The results of $\overline{\eta}$, $f_{\text{Im}}$, $\epsilon$, and $w_h$ as functions of $V_1$ and $V_2$, respectively. (d) The energy spectra for the three symbols (triangle, square and circle) in (b), with $V_1=0.5, 1, 1.5$ and $V_2=0.5$, respectively. (f) $\bar{\eta}$ (blue dash-dotted line), $|w_h|$ (red dotted line), $f_{\text{Im}}$ (black dashed line), and $\bar{\epsilon}$ (cyan solid line) as functions of $V_1$ with fixed $V_2=0.5$. Other parameters are $g=0$ and $h=0.5$.
	}\label{fig3}
\end{figure*}

We first study the nonreciprocal effect on the localization phase diagram showed in Fig. \ref{fig1}(b).
The three different phases can be distinguished from the localization properties indicated by the averaged fractal dimension (FD) over all eigenstates of the model Hamiltonian:
\begin{equation}
\overline{\eta}=\frac{1}{D}\sum_{n=1}^{D}\eta_n=-\frac{1}{D}\sum_{n=1}^{D}\frac{\ln(\sum_j^D|\braket{j|n}|^4)}{\ln D},
\end{equation}
where $\ket{n}$ and $\ket{j}$ denotes the $n$-th right eigenstates and computational basis respectively, and $D=\binom{L}{N}$ is the total number of the eigenstates and thus the dimension of the Hilbert space.
Note that $\sum_j^D|\braket{j|n}|^4=\sum_j^D|\psi_n(j)|^4$ is the inverse participation ratio, which is widely used in the single particle and interacting systems \cite{LZTang2020,GQZhang2020}. We display the averaged FD $\overline{\eta}$ in the parameter space $V_1$-$V_2$ in Fig.~\ref{fig2}(a) for typical parameters $g=0.5$ and $h=0$. The extended, localized, and critical phases can be clearly recognized with $\overline{\eta}\to1$, $\overline{\eta}\to0$, and $0<\overline{\eta}<1$, respectively.
As shown in Fig. \ref{fig1}(b) and Fig.~\ref{fig2}(a), one can find that the non-reciprocity can enhance the delocalized phase (i.e., the extended and critical phases), whereas the boundary between the extended and localized phases is also modified by the non-reciprocity. We find that the critical value of localization-delocalization transition $V_{1c}$ can be described by the equation:
\begin{equation}\label{boundary}
V_{1c}=e^{-|h|}\left(2K\cosh|g|+2\sqrt{K^2-V_2^2}\sinh|g|\right)
\end{equation}
with $K=\max(t,V_2)$, and is plotted as the red dashed line in Fig.~\ref{fig2}(a). Despite the lacking of an analytical proof, the above relation works well in separating the localized and delocalized phases in different non-Hermitian cases [see also Fig.~\ref{fig3}(a) with $g=0, h=0.5$ and Fig.~\ref{fig4}(a) with $g=h=0.5$].  Notably, Eq.~(\ref{boundary}) can be considered as an empirical combination of the corresponding analytical results under three limitations. In the Hermitian limit with $g=h=0$, Eq.~(\ref{boundary}) reduces to $V_{1c}=2\max(t,V_2)$, which is analytically derived in Ref. \cite{FLLiu2014}. When $V_2=0$ and in the reciprocal hopping ($g=0$) and real potential ($h=0$) limits, Eq.~(\ref{boundary}) reduces to $V_{1c}=2te^{-|h|}$ as derived in Ref.~\cite{Longhi2019a} and $V_{1c}=2te^{|g|}$ as derived in Ref.~\cite{HJiang2019}, respectively.
When $V_2=0$, this model reduces to the nonreciprocal AAH model and the transition boundary given by Eq. (\ref{boundary}) is simplified to a transition point $V_{1c}=2e^{|g|}\approx3.2974$ with $g=0.5$~\cite{HJiang2019}.
Interestingly, when $2\cosh|g|<V_1<2e^{|g|}$, there is a phase transition from the extended to localized and to critical phase can be induced by increasing $V_2$. Such a phase transition is absent in the Hermitian case. Figure \ref{fig2}(b) displays the typical density distributions $|\psi_n(j)|^2$ of the ground state ($n=1$) for the three different phases. The corresponding averaged FD $\bar{\eta}$ as a function of the system size $L$ are shown in Fig. \ref{fig2}(c). The results show that the localized, extended, and critical phases can be well identified by the density distribution and the averaged FD in non-Hermitian systems. Notably, there is no mobility edge in our model for the noninteracting case and thus all states share the same localization behavior.

We turn to study the eigenenergy spectrum and the related winding numbers defined solely for non-Hermitian systems. One can define the ratio of the complex eigenenergies in the whole spectrum as
\begin{eqnarray}
f_{\text{Im}}={D_{\text{Im}} / D},
\end{eqnarray}
where $D_{\text{Im}}$ is the number of eigenenergies whose imaginary part $\mathrm{abs}(\mathrm{Im}(E))>\textit{C}$, with the cutoff of imaginary part $\textit{C}=10^{-13}$ is considered as the error introduced in our numerical diagonalization. Figure \ref{fig2}(d) shows the numerical results of $f_{\text{Im}}$ as a function of $V_1$ and $V_2$. We find that the whole energy spectrum is either real ($f_{\text{Im}}=0$) or complex ($f_{\text{Im}}=1$) in the phase diagram and the boundaries of the real-complex transition coincides with the red dashed line defined by Eq. (\ref{boundary}). This indicates that the real-complex transition is accompanied by the localization-delocalization transition. In this case, the localized phase corresponds to a parity-time-symmetry preserved phase, while the extended and critical phases correspond to the parity-time broken phase with the eigenenergy spectrum forms several rings encircling different base points in the complex energy plane, as shown in Fig. ~\ref{fig2}(e).

One can use the following winding number to characterize the topology of the complex eigenenergies ~\cite{ZGong2018,Longhi2019a,Longhi2019b,HJiang2019}
\begin{equation}\label{winding number}
w_{\nu}=\int_{0}^{2 \pi} \frac{d \theta_{\nu}}{2 \pi i} \partial_{\theta_{\nu}}\ln \operatorname{det}\left[H\left(\theta_{\nu}\right)-E_{B}\right]
\end{equation}
with $\nu=g,h$, where $E_B$ is the energy base not belonging to the spectrum. Since the distribution of the spectrum behaves more complex than a Cantor set in this case, the base energy $E_B$ needs a redefinition for different $V_1$ and $V_2$ rather than simply setting $E_B=0$~\cite{Longhi2019}. Unlike the conventional winding numbers for eigenstates which response to the bulk-edge correspondence, here $w_\nu$ counts how many times the complex eigenenergy trails encircle the energy base $E_B$ when varying the phase $\theta_\nu$ from $0$ to $2\pi$. Thus, the winding number $w_\nu$ is not related to topological edge states and different values of $w_\nu$ just imply the topological phase transition. Recently, it was revealed that the values of $w_\nu$ determine the number of skin bulk modes in the non-Hermitian skin effect when the non-Hermitian lattice systems are under open boundary conditions~\cite{Zhang2020,Yang2020,Okuma2020}. For complex energy spectra, the eigenenergies belonging to each ring move alone their own circular path as $\theta_{\nu}$ varies form $0$ to $2\pi$, thus we can get a non-trivial winding number from Eq. (\ref{winding number}) by choosing a proper $E_B$ near the center of each ring. For real spectra, the eigenenergies locate at the real axis when varying $\theta_{\nu}$, and the corresponding winding number $w_{\nu}=0$ is independent of $E_B$. The eigenenergies remain unchanged when varying $\theta_{h}$ in the nonreciprocal hopping case, and the calculated $w_g$ as a function of $V_1$ and $V_2$ is shown in Fig.~\ref{fig2}(f). The result reveals that the topological phase transition is accompanied with the localization transition given by the red dashed line. Thus, the single-particle localization transition, the real-complex transition, and the topological phase transition happen simultaneously in the nonreciprocal quasiperiodic lattice.

\subsection{Complex potential phase case}

We now consider the on-site potential with the complex phase $h\neq0$. Typical results for $h=0.5$ and $g=0$ are shown in Fig. \ref{fig3}. Comparing the results of $\overline{\eta}$ [Fig.~\ref{fig3}(a)] with those in the Hermitian limit [Fig.~\ref{fig1}(b)], we find that the delocalization-localization transition happens at smaller $V_1$. Thus the complex phase can enlarge the localization region, which is opposite to enlargement of the delocalization region by the nonreciprocal hopping. The effect of the complex phase can be viewed as a renormalization of $V_1$ with an effective value $\tilde{V_1}=V_1e^{|h|}$, with the phase boundary governed by Eq. (\ref{boundary}) with $g=0$.

\begin{figure}[tb]
	\centering
	\includegraphics[width=0.48\textwidth]{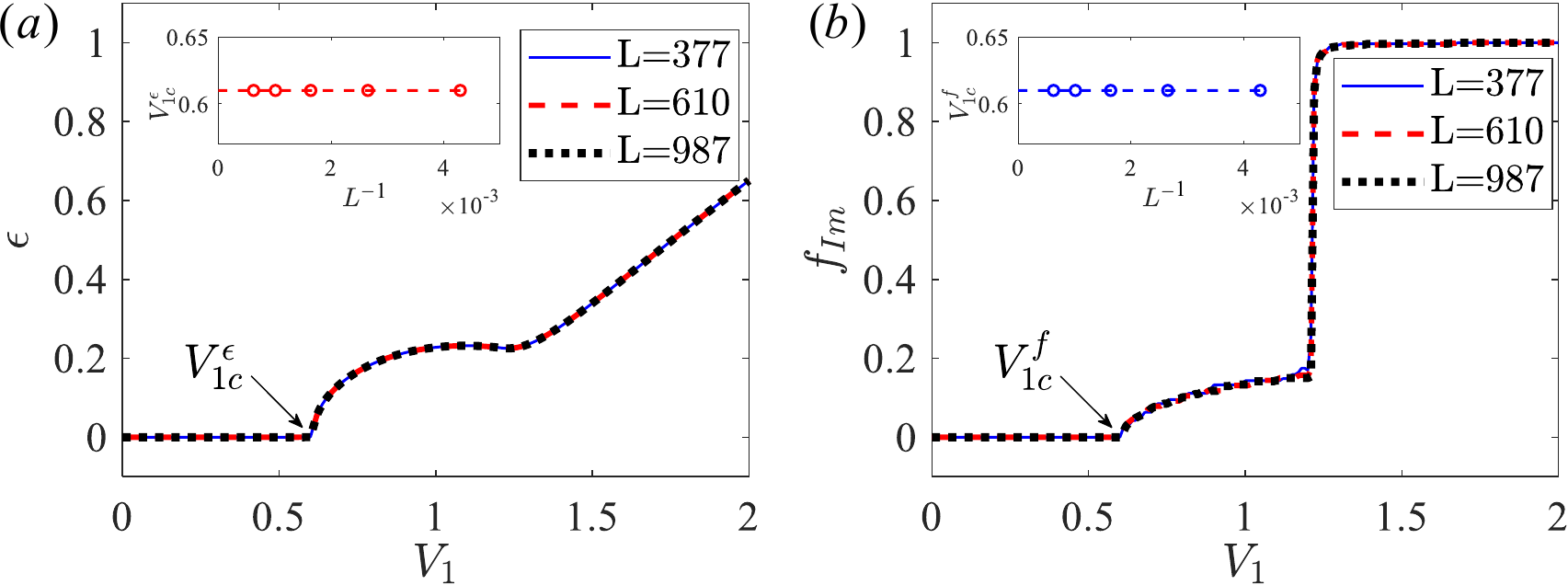}
	\caption{(Color online) (a) The ratio of the complex eigenenergies $f_{\mathrm{Im}}$ and (b) the largest imaginary part of the spectrum $\epsilon$ as a function of $V_1$. The results for $L=377$, 610 and 987 are plotted as blue solid, red dashed, and black dotted lines, respectively. The real-complex transition points $V_{1c}^f$ and $V_{1c}^\epsilon$ as a function of $1/L$ are displayed in the inset plots. Other parameters are the same with those in Fig.~\ref{fig3} (f).
	}\label{fig4}
\end{figure}

In order to study the real-complex transition of the energy spectrum, we calculate the ratio $f_{\text{Im}}$ and the largest imaginary part in the whole spectrum $\epsilon$ as functions of $V_1$ and $V_2$, with the results shown in Figs. \ref{fig3}(b) and \ref{fig3}(c), respectively. In sharp contract to the nonreciprocal hopping case, we find $f_{\text{Im}}\approx1$ and $\epsilon\neq0$ in the localized phase, which indicates that all the localized states have complex energies. However, the extended states can have real ($f_{\text{Im}}=0$ and $\epsilon=0$) or partially complex energies in this case, and the real-complex transition happens in the extended phase. Three typical energy spectra in the complex plane when $\theta_{\nu}$ varies from $0$ to $2\pi$ and the corresponding winding numbers $w_h$ are plotted in Fig. \ref{fig3}(d). One can see that a trivial patten with $w_h=0$ can form in the complex energy plane when varying $\theta_{\nu}$. We further calculate $w_h$ as functions of $V_1$ and $V_2$, as shown in Fig. \ref{fig3}(e). In this complex potential case, the topological phase transition between $w_h=1$ and $w_h=0$ is always consistent with the localization-delocalization transition, which is the same to the nonreciprocal hopping case. When $V_2=0$, these two phase transitions are consistent with the real-complex transition~\cite{Longhi2019}, with the critical value $V_{1c}=2e^{-0.5}\approx1.2131$. However, when $V_2\neq0$, there is no such coincidence and the real-complex transition can happen before the localization and topological transitions when increasing $V_1$ with fixed $V_2$, as depicted in Fig. \ref{fig3}(f). We also study the finite-size scaling of the real-complex transition. The ratio of the complex eigenenergies $f_{\mathrm{Im}}$ and the largest imaginary part of the spectrum $\epsilon$ for the lattice sizes $L=377,~610,~987$ (chosen from the Fibonacci sequences) are shown in Figs.~\ref{fig4}(a) and \ref{fig4}(b), respectively. The inset plots show the convergence of the real-complex transition points as increasing $L$. These results demonstrate that the real-complex transition is robust against the finite-size effect.

\subsection{Coexistence of two non-Hermiticities}
\begin{figure}[tb]
	\centering
	\includegraphics[width=0.45\textwidth]{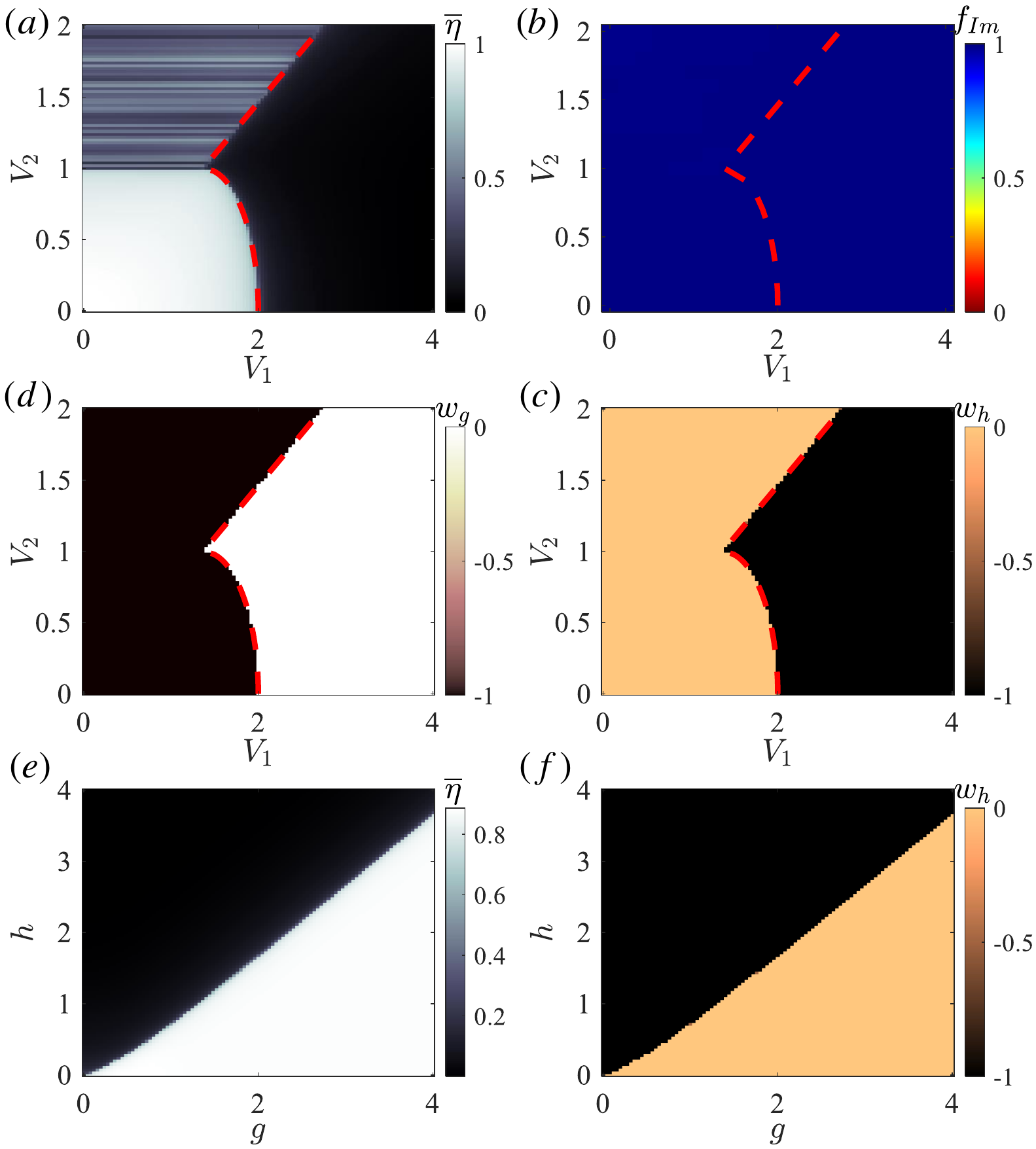}
	\caption{(Color online) Generalized AAH model with both nonreciprocal hopping and complex potential phase for the system size $L=610$. (a-d) $\overline{\eta}$, $f_{\text{Im}}$, $w_g$, and $w_h$ as functions of $V_1$ and $V_2$ for $g=0.5$ and $h=0.5$, respectively. (e,f) $\bar{\eta}$ and $w_h$ as functions of $g$ and $h$ for $V_2=2V_1=2$, respectively.
	}\label{fig5}
\end{figure}

Finally in this section, we consider the coexistence of nonreciprocity and complex potential phase in our model.
The averaged FD $\overline{\eta}$ and the ratio $f_{\text{Im}}$ as functions of $V_1$ and $V_2$ for $g=h=0.5$ are plotted in Figs.~\ref{fig5}(a) and \ref{fig5}(b), respectively.
Since all the eigenenergy spectrums in the delocalized (localized) phase are complex for nonreciprocal hopping (complex potential phase) cases [see Figs.~\ref{fig2}(d) and \ref{fig3}(b)], the combination of nonreciprocity and complex phase brings complex eigenenergies for the localized and delocalized states, as shown in Fig. \ref{fig5}(b). Thus, there is no real-complex transition in the quasiperiodic lattice under the coexistence of nonreciprocity and complex potential phase. Figures~\ref{fig5}(c) and \ref{fig5}(d) show the winding numbers $w_g$ and $w_h$ as functions of $V_1$ and $V_2$, both of which describes the boundary of the topological phase transition in the parameter space and the topological nature of the localization-delocalization transition [the red dashed line given by Eq. (\ref{boundary})]. Moreover, we plot $\overline{\eta}$ and $w_h$ as functions of $g$ and $h$ for $V_1=2$ and $V_2=0.9$ in Figs. \ref{fig5}(e) and \ref{fig5}(f), which show the competition between these two kinds of non-Hermiticities in the localization and topological transitions.

\section{\label{sec4} results in interacting cases}
We proceed to study the non-Hermitian MBL in the quasiperiodic lattice in the presence of interactions. In our simulations based on the exact diagonalization, we take the interaction strength $U=2$ and the half filling condition (the particle number of the spinless fermions $N_a=L/2$) in the lattice with the size up to $L=14$. To reduce the statistical error for relatively small lattices in the exact diagonalization of the many-body Hamiltonian in Eq. (\ref{H}), we add a phase shift $\phi$ randomly chosen from $0$ to $2\pi$ into the hopping and potential modulation terms in Eq. (\ref{H2}): $\theta_h/L\rightarrow\theta_h/L+\phi$. The numerical results in this section are averaged over the random phase $\phi$ with sufficient samples.

\begin{figure}[tb]
	\centering
	\includegraphics[width=0.48\textwidth]{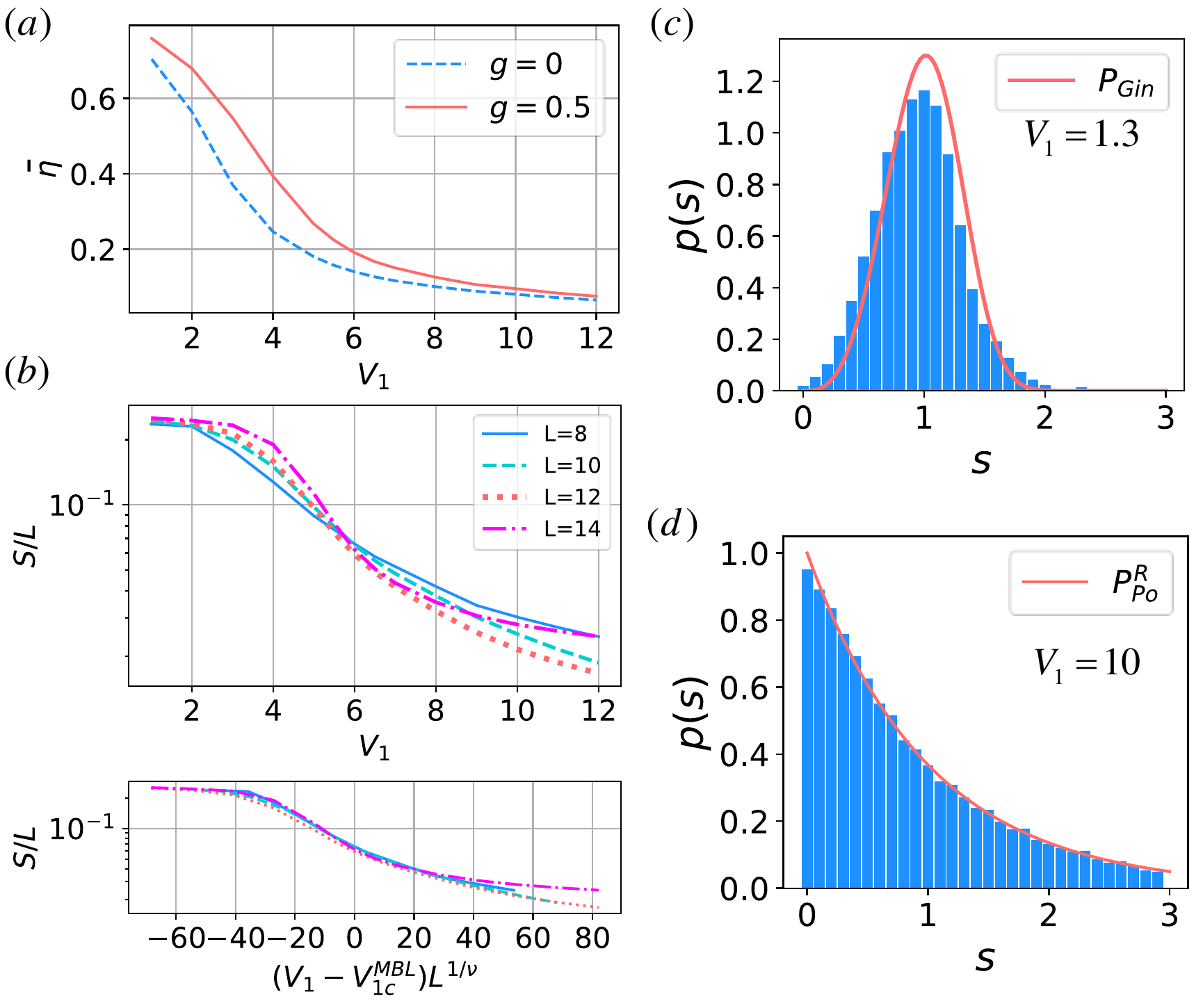}
	\caption{(Color online) MBL transition in the interacting nonreciprocal generalized AAH model. (a) The averaged FD $\overline{\eta}$ as a function of $V_1$ for $g=0,0.5$. $\overline{\eta}$ is averaged over the mid-$1/6$ of the real part in the spectrum and the system size $L\in\{8,10,12,14\}$. (b) (top) Averaged entanglement entropy per lattice site $S/L$ as a function of $V_1$ for different $L$. $S/L$ is averaged over the center $1/10$ of the whole spectrum in the complex plane. (bottom) Critical scaling collapse of $S/L$ as a function of $(V_1-V_{1c}^{\text{MBL}})L^{1/\nu}$ with $V_{1c}^{\text{MBL}}=6.0$ and $\nu=2$. (c, d) The probability distribution of nearest-level-spacing $p(s)$ for $V_1=1.3$ and $V_1=10$ with $L=12$, respectively. Other parameters are $g=0.5,~h=0,~V_2=0.5$, and $U=2$.
	}\label{fig6}
\end{figure}

\begin{figure}[tb]
	\centering
	\includegraphics[width=0.48\textwidth]{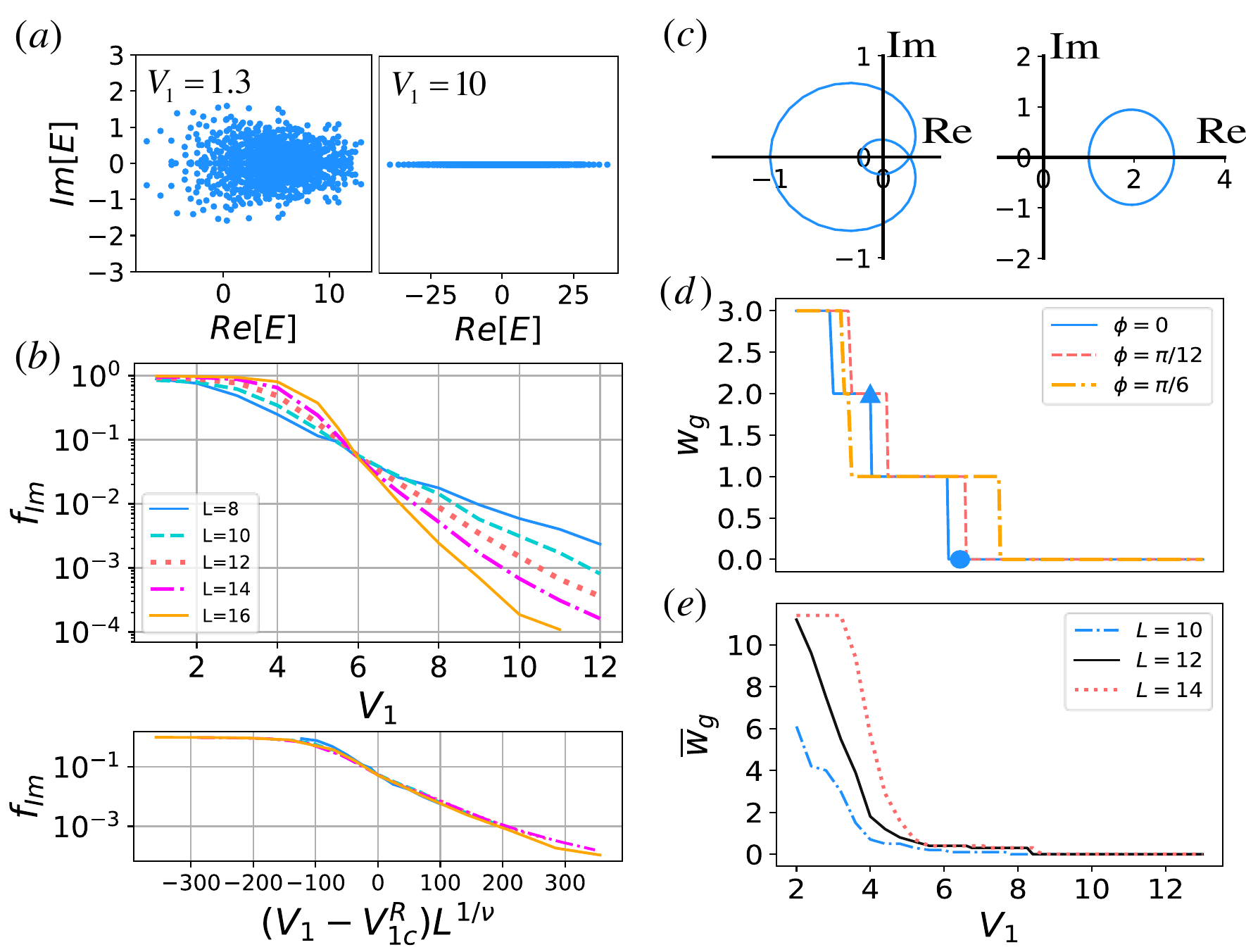}
	\caption{(Color online) Real-complex and topological phase transition in the interacting nonreciprocal generalized AAH model. (a) Real-complex transition of eigenenergy spectra for $V_1=1.3$ (left) and $V_1=10$ (right) for the phase shift $\phi=0$. (b) (top) $f_{\text{Im}}$ as a function of $V_1$ for different $L$. (bottom) Critical scaling collapse of $f_{\text{Im}}$ as a function of $(V_1-V_{1c}^{R})L^{1/\nu}$ with $V_{1c}^{R}=6.0$ and $\nu=0.7$. (c) The $\theta_g$ dependence of $\det H(\theta_g)/|\det H(0)|$ with $\theta_g$ varies from $0$ to $2\pi$ for $\phi=0$, $V_1$=4 (left) and $V_1=6.43$ (right), corresponding to the triangle and round symbols in (d). Here $w_g$ counts the times that the blue curve encircles around the origin $E_B=0$.(d) $w_g$ as a function of $V_1$ for different $\phi$ and $L=10$. (e) The averaged winding number $\overline{w}_g$ (over different samples of $\phi$) as a function of $V_1$ for $L=10,12,14$. Other parameters are $g=0.5$, $h=0$, $V_2=0.5$, and $U=2$.
	}\label{fig7}
\end{figure}

We first consider the nonreciprocal case with typical parameters $g=0.5$, $h=0$, and $V_2=0.5$.
To catch a glimpse of the nonreciprocal effect in the MBL, we calculate the averaged FD $\overline{\eta}$ as a function of $V_1$ for $g=0,0.5$ in Fig. \ref{fig6}(a). One can see that the nonreciprocal hopping enlarges the $\overline{\eta}$.
In order to determine the MBL transition, we investigate the scaling of the half-chain EE for different system size $L$, which have been utilized to identify the extended phase from the MBL phase in non-Hermitian systems~\cite{LJZhai2020,Hamazaki2019}.
For the $n$-th normalized right eigenstate $\ket{\psi_n}$, the half-chain EE is defined as $S_n=-\mathrm{Tr}\rho_n\ln\rho_n$, where $\rho_n=\mathrm{Tr}_{L/2}\ket{\psi_n}\bra{\psi_n}$ is the reduced density matrix.
We average the EE over $1/10$ eigenstates from those energy locating near the center of the spectrum in the complex plane and take sufficient phase shift samples. The averaged EE per lattice site $S/L$ as a function of $V_1$ for different $L$ and the critical scaling collapse of $S/L$ are plotted in Fig. \ref{fig6}(b). We find that $S/L$ subjects to the volume law when $V_1$ is small in the extended phase, and the crossover of averaged EE for different $L$ indicates the breakdown of the volume law, and there is a MBL transition occurring at $V_1=V_{1c}^{\text{MBL}}\approx6.0$ obtained from the critical scaling fitting.

Furthermore, we study the level statistics of the complex energy spectrum by considering the nearest-level-spacing $s_n\equiv\min\{|E_n-E_m|\}$, which denotes the minimum distance between two eigenenergies $E_n$ and $E_m$ in the complex plane.
The probability distribution of $\{s_n\}$ denoted by $p(s)$ is plotted in Figs. \ref{fig6}(c) and \ref{fig6}(d) for $V_1=1.3$ and $10$, respectively. For $V_1=1.3$, the system is in the extended phase and the distribution $p(s)$ approaches to the Ginibre distribution $P_{\mathrm{Gin}}^{\mathrm{C}}(s)=cp(cs)$, which characterizes an ensemble for non-Hermitian Gaussian random matrices~\cite{Hamazaki2019}:
\begin{equation}
p(s)=\lim _{N \rightarrow \infty}\left[\prod_{n=1}^{N-1} e_{n}\left(s^{2}\right) e^{-s^{2}}\right] \sum_{n=1}^{N-1} \frac{2 s^{2 n+1}}{n ! e_{n}\left(s^{2}\right)}
\end{equation}
with
\begin{equation}
e_{n}(x)=\sum_{m=0}^{n} \frac{x^{m}}{m!} ~~ \text{and} ~~c=\int_{0}^{\infty}dssp(s)=1.1429 \cdots \nonumber.
\end{equation}
For $V_1=10$, the system is in the MBL phase, and the eigenenergies fall to the real axis in this case and thus obey the (real) Poisson distribution $P_{\mathrm{Po}}^{R}(s)=e^{-s}$. These results demonstrate that the MBL phase can be distinguished from the extended phase by the averaged EE and the level statistics.

We turn to study the real-complex transition and the topological phase transition in the system. In Fig.~\ref{fig7}(a), we plot the eigenenergy spectra for $V_1=1.3$ and $10$, where the spectra are symmetric with respect to the real axis due to the time-reversal symmetry and the imaginary part of the spectrum are suppressed when $V_1$ becomes large. The ratio
$f_{\text{Im}}$ as a function of $V_1$ for different $L$ is plotted in Fig.~\ref{fig7}(b). We find that $f_{\text{Im}}$ increases (decreases) when $V_{1}\lesssim6.0$ ($V_{1}\gtrsim6.0$) with real-complex transition occurring at $V_1=V_{1c}^R\approx6.0$, which is obtained from the critical scaling collapse of $f_{\text{Im}}$. Thus, we conjecture that the MBL transition is accompanied by the real-complex transition with $V_{1c}^R=V_{1c}^{\text{MBL}}$ in this nonreciprocal generalized AAH model of interacting spinless fermions.

To determine the energy winding number in this case, we plot the $\theta_g$-dependence of $\det H(\theta_g)/|\det H(0)|$ in Fig.~\ref{fig7}(c) to illustrate the loop of complex energy winding around the origin $E_B=0$. The times of the loop winding around the origin gives the value of $w_g$ in Eq. ({\ref{winding number}}). We find that the obtained $\theta_g$ depends on the phase shift $\phi$, and $w_g$ as a function of $V_1$ for different $\phi$ is plotted in Fig.~\ref{fig7}(d). The averaged winding number $\overline{w}_g$ over different samples of $\phi$ for $L=10$ is shown in Fig.~\ref{fig7}(e), which indicates a topological phase transition between the phases $\overline{w}_g>0$ (nonzero integers of $w_g$ for some values of $\phi$) and $\overline{w}_g=0$ ($w_g=0$ for all $\phi$) at $V_1=V_{1c}^{T}\approx7.8>V_{1c}^{\text{MBL}}$. This indicates that when $6.5<V_1<7.8$, the interacting system of $L=10$ is in the MBL phase with nonzero eigenenergy winding numbers. As shown in Fig.~\ref{fig7}(e), we also calculate the topological phase transition for the lattice size $L=12$ and $14$ with the obtained critical points $V_{1c}^{T}\approx8.6$ and 8.7, respectively. Thus, the shift of the topological phase transition from the MBL transition tends to increase and then becomes robust as increasing $L$ from 10 to 14. These results show that the many-body non-Hermitian interacting systems have more complicated topological characters than the single-particle counterparts.


\begin{figure}[tb]
	\centering
	\includegraphics[width=0.48\textwidth]{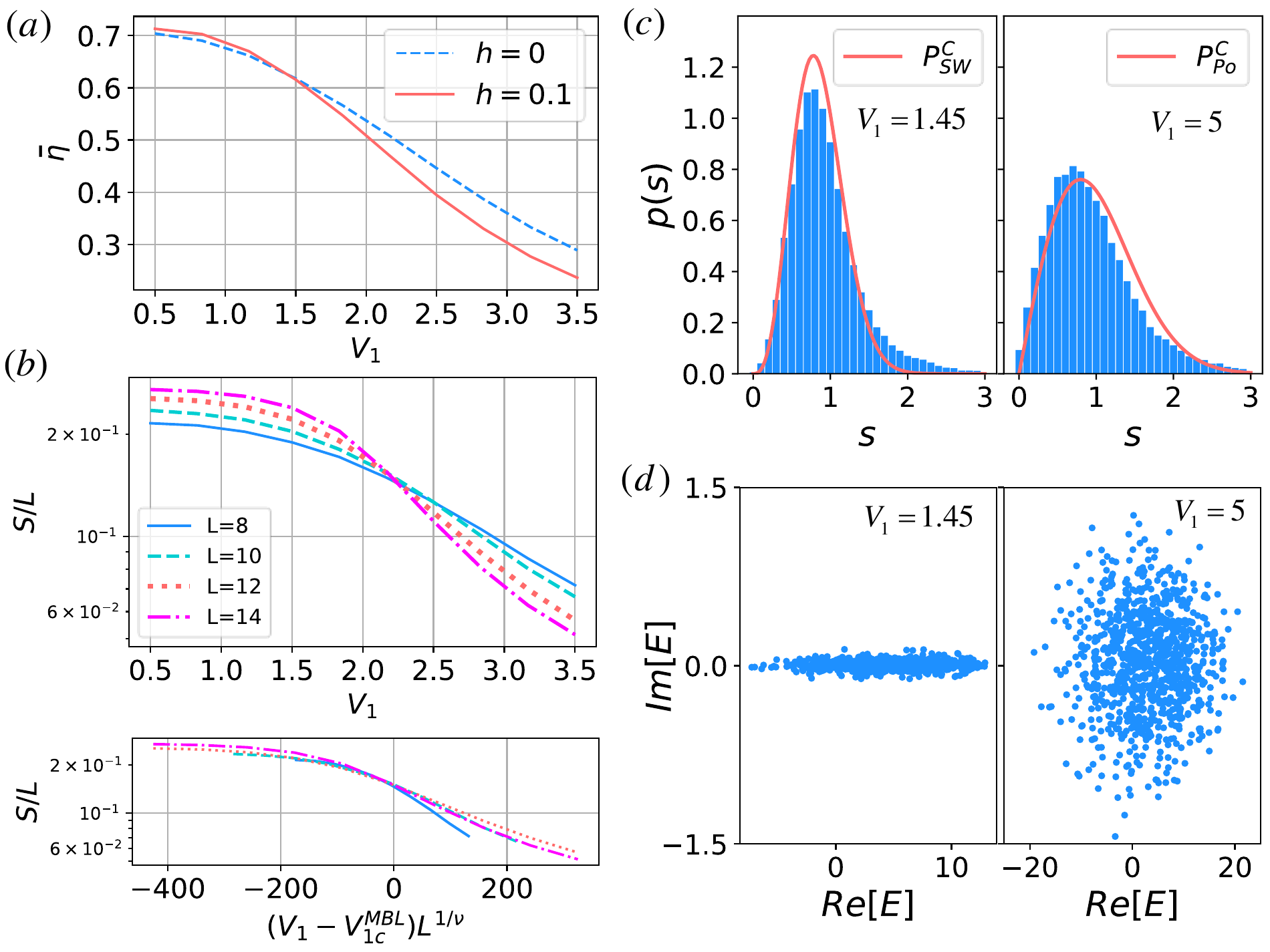}
	\caption{(Color online) MBL in the interacting generalized AAH model with the complex potential phase. (a) $\overline{\eta}$ averaged over all eigenstates as a function of $V_1$ for $h=0,~0.1$. (b) (top) $S/L$ as a function of $V_1$ for different $L$. (bottom) Critical scaling collapse of $S/L$ as a function of $(V_1-V_{1c}^{\text{MBL}})L^{1/\nu}$ with $V_{1c}^{\text{MBL}}=2.2$ and $\nu=0.45$. (c) $p(s)$ for $V_1=1.45$ (left) and $V_1=5$ (right) with $L=12$. (d) Energy spectra for $V_1=1.45$ (left) and $V_2=5$ (right) with $L=12$ and $\phi=0$. Other parameters are $g=0$, $h=0.1$, $V_2=0.5$, and $U=2$.
	}\label{fig8}
\end{figure}

We further study the MBL phase in the complex potential case. The averaged FD $\overline{\eta}$ as a function of $V_1$ for $g=0$, $h=0.1$ and $V_2=0.5$ is plotted in Fig.~\ref{fig8}(a). To determine the MBL transition, we plot the averaged EE per lattice site $S/L$ as a function of $V_1$ for different $L$ and the corresponding critical scaling collapse of $S/L$ in Fig.~\ref{fig8}(b). One can find MBL transition at $V_1=V_{1c}^{\text{MBL}}\approx2.2$ in this case. We also calculate the probability distribution $p(s)$ of the level spacing in the complex energy space, with the results for $V_1=1.45$ and $V_1=5$ shown in Fig.~\ref{fig8}(c), which corresponds to the extended and MBL phases, respectively. For the non-Hermitian extended phase ($V_1=1.45$), $p(s)$ approach to sub-Wigner probability distribution~\cite{Tzortzakakis2020}: $P_{\text{SW}}(s)=as^{b}e^{-cs^{2}}$ with the fitting parameters $b=2.8$ and $c=2.3$, while $p(s)$ approaches to complex Poisson distribution~\cite{Hamazaki2019}: $P_{\mathrm{Po}}^{\mathrm{C}}(s)=\pi s/2e^{-(\pi/4)s^{2}}$ in the MBL phase ($V_1=5$). Therefore, the non-Hermitian MBL phase has different level statistics in the nonreciprocal hopping (real Poisson) and complex potential (complex Poisson) cases. In addition, we find that the real-complex  transition and the topological phase transition are absent in this interacting generalized AAH model with the complex potential phase since the eigenenergies are generally complex. Typical complex energy spectra in the extended and MBL phases are shown in Fig.~\ref{fig8}(d).

\section{SUMMARY AND OUTLOOK}\label{sec5}

In summary, we have explored the localization (MBL) transition and topological phase transition in the (interacting) non-Hermitian generalized AAH model with the non-Hermiticities from the nonreciprocal hopping and complex potential phase. In the single-particle regime, we reveal that the nonreciprocal hopping (the complex potential phase) can enlarge the delocalization (localization) region in the phase diagrams, and the localization transition are always accompanied by the topological phase transition in different non-Hermitian cases. Moreover, we found that a real-complex transition in the energy spectrum coincides with (occurs before) these two phase transitions in the nonreciprocal (complex potential) case, while the real-complex transition is absent under the coexistence of the two non-Hermiticities. For interacting spinless fermions, we have shown that the extended phase and the MBL phase can be identified by the entanglement entropy and different level statistics. We also found that the MBL transition coincides with the real-complex transition and occurs before the topological phase transition in the nonreciprocal case, which are absent in the complex phase case.

Finally, we make some remarks on the future research. The interplay among the localization, interaction, and non-Hermitian skin effects under open boundary conditions would be an interesting research topic. It would be also valuable to further investigate the non-Hermitian MBL phase with topologically nontrivial characters. The non-Hermitian critical phase in the presence or absence of interactions in our model may have some exotic localization properties, such as the non-Hermitian critical statistics and the phase transitions. Furthermore, the non-Hermitian effects on the localization and topological properties in other types of generalized AAH models can be studied, such as the generalized AAH models with non-Abelian or long-range hopping terms.

\begin{acknowledgments}
We thank L.-J. Lang and S.-L. Zhu for helpful discussions. This work was supported by the National Natural Science
Foundation of China (Grants No. U1830111 and No. 12047522), the Key-Area Research and Development Program of Guangdong Province (Grant No. 2019B030330001), the Science and Technology of Guangzhou (Grants No. 2019050001 and No. 201804020055), and the Guangdong Basic and Applied Basic Research Foundation (Grant No. 2020A1515110290).
\end{acknowledgments}

\bibliography{reference}

\end{document}